\documentclass[aps,prl,reprint,groupedaddress]{revtex4-1}
\usepackage{amssymb}
\usepackage{amsmath}
\usepackage{array}
\usepackage{multirow,bigdelim}
\usepackage{enumitem}
\usepackage{xcolor}

\begin{document}

\title{Comparing the three W-like states with the W state}
\author{ Dafa Li}

\affiliation{Department of Mathematical Sciences, Tsinghua University, Beijing 100084, China\\ The corresponding author is Dafa Li\\ email is lidafa@tsinghua.edu.cn}

\begin{abstract}
Abstract: In [Phy. Rev. Lett., 98, 260501 (2007)] and previous papers, the
W-like states were used in many quantum communication schemes, distillation,
teleportation and superdense coding. So far, no one has compared the W-like
states with the W state. We investigate the properties of three W-like
states named as $|\vartheta ^{\prime }\rangle $, $|\eta ^{\prime }\rangle $
and $|\xi ^{\prime }\rangle $. We point out that $|\vartheta ^{\prime
}\rangle $ (resp. $|\eta ^{\prime }\rangle $ and $|\xi ^{\prime }\rangle $)
has the maximal von Neumann entanglement entropy (vNEE) $S(\rho _{A})=\ln 2$
(resp. $S(\rho _{B})=\ln 2$ and $S(\rho _{C})=\ln 2$) and the maximal tangle
$\tau _{A(BC)}=1$ (resp. $\tau _{B(AC)}=1$ and $\tau _{C(AB)}=1$) while the
W state does not. We also indicate that if some one of three qubits is
traced out, then the remaining two qubits of $|\vartheta ^{\prime }\rangle $
(resp. $|\eta ^{\prime }\rangle $ and $|\xi ^{\prime }\rangle $) are more
entangled than the two qubits of the W state by several entanglement
measures. It means that the three W-like states have higher robustness
against some particle loss than the W state. We show that $|\vartheta
^{\prime }\rangle $, $|\eta ^{\prime }\rangle $ and $|\xi ^{\prime }\rangle $
can be suitable for perfect teleportation and superdense coding but the W
state cannot.

\end{abstract}

\maketitle

\section{Introduction}

D\"{u}r et al. partitioned pure states of three qubits into six stochastic
local operations and classical communication (SLOCC)\ equivalence classes,
two of which are the SLOCC\ classes GHZ and W, which are genuinely entangled
\cite{Dur}. The main difference between the GHZ\ and W states is that if one
of the three qubits is traced out, then the state of the remaining 2-qubit
system is still entangled for the W state while separable for the GHZ state.
A state of three qubits $A,B,$and $C$\ is referred to as the maximal
entangled state if\ the reduced density matrices obtained by tracing out of
any two qubits are proportional to the identity (i.e. $S(\rho _{x})=\ln 2$, $%
x=A,B,$and $C$) \cite{Gour}. By the definition, the GHZ state is the unique
maximally entangled state for three qubits under local unitary operators
(LU) \cite{Li-qip, Li-qic}. We found the states of the GHZ SLOCC\ class with
the maximal vNEE $S(\rho _{x})=\ln 2$, where $x=A,B,$or $C$ in \cite{Li-qic}%
. By using the binary matrix and alternative matrix, Osterloh and Siewert
explored the maximally entangled states defined in their Definition II.1%
\textbf{\ }\cite{Osterloh}, and by using the comb and filters, they
distinguished the maximally entangled states \cite{Osterloh-ijqi}.

The states of the form $\gamma |001\rangle +\beta |010\rangle +\alpha
|100\rangle $, where $\alpha \beta \gamma \neq 0$, are called the W-like
states and the properties of the W-like states were discussed in the
previous papers. The W-like states are more general quantum entanglement
states used in many quantum communication schemes, distillation,
teleportation, and superdense coding \cite{Lo, Yang, Aug, Ding, Ke,
Fer,Dong, Wu}. But, so far, no one has compared the W-like states with the W
state.

We explore the following three W-like states $|\vartheta ^{\prime }\rangle $%
, $|\eta ^{\prime }\rangle $and $|\xi ^{\prime }\rangle $, any two of which
are inequivalent under LU \ and all of which are LU\ inequivalent to the W
state, and compare $|\vartheta ^{\prime }\rangle $, $|\eta ^{\prime }\rangle
$ and $|\xi ^{\prime }\rangle $ with the W state via four entanglement
measures: the 2-tangle $\tau _{xy}$, the negativity, vNEE, and the tangle $%
\tau _{x(yz)}$. We show that $|\vartheta ^{\prime }\rangle $, $|\eta
^{\prime }\rangle $ and $|\xi ^{\prime }\rangle $ can be suitable for
perfect teleportation and superdense coding but the W state cannot.

\begin{eqnarray*}
|\vartheta ^{\prime }\rangle &=&\frac{1}{2}|001\rangle +\frac{1}{2}%
|010\rangle +\frac{1}{\sqrt{2}}|100\rangle \\
|\eta ^{\prime }\rangle &=&\frac{1}{2}|001\rangle +\frac{1}{\sqrt{2}}%
|010\rangle +\frac{1}{2}|100\rangle \\
|\xi ^{\prime }\rangle &=&\frac{1}{\sqrt{2}}|001\rangle +\frac{1}{2}%
|010\rangle +\frac{1}{2}|100\rangle
\end{eqnarray*}

\section{Preliminary}

It is known that each state of the W SLOCC\ class can be transformed into
the following Schmidt decomposition (SD) form under LU \cite{Acin00,
Dli-qip-18, Dli-jpa-20}

\begin{equation}
|\psi \rangle =\lambda _{0}|000\rangle +\lambda _{1}e^{i\phi }|100\rangle
+\lambda _{2}|101\rangle +\lambda _{3}|110\rangle ,  \label{w-0}
\end{equation}%
where $\lambda _{i}>0$, $i=0,2,3$, $\lambda _{1}\geq 0$, $%
\sum_{i=0}^{3}\lambda _{i}^{2}=1$, and $0\leq \phi <2\pi $ and $\phi $ is
called a phase.

For\ the W state $\frac{1}{\sqrt{3}}(|001\rangle +|010\rangle +|100\rangle )$%
, its SD is $\frac{1}{\sqrt{3}}(|000\rangle +|101\rangle +|110\rangle )$.
The following states are also referred to as the W-like states.
\begin{equation}
\lambda _{0}|000\rangle +\lambda _{2}|101\rangle +\lambda _{3}|110\rangle ,
\label{w-like}
\end{equation}%
where $\lambda _{i}>0$, $i=0,2,3$, and $\lambda _{3}^{2}+\lambda
_{2}^{2}+\lambda _{0}^{2}=1$. Note that the W-like state $\gamma |001\rangle
+\beta |010\rangle +\alpha |100\rangle $ is LU equivalent to its SD $|\alpha
||000\rangle +|\gamma ||101\rangle +|\beta ||110\rangle $.

It is easy to compute the 2-tangle $\tau _{xy}$, the negativity, vNEE, and
the tangle $\tau _{x(yz)}$ via SD of a state. For the three W-like states $%
|\vartheta ^{\prime }\rangle $, $|\eta ^{\prime }\rangle $, and $|\xi
^{\prime }\rangle $, a calculation yields the following SDs: $|\vartheta
\rangle $, $|\eta \rangle $, and $|\xi \rangle $, respectively.

\begin{eqnarray*}
|\vartheta \rangle &=&\frac{1}{\sqrt{2}}|000\rangle +\frac{1}{2}|101\rangle +%
\frac{1}{2}|110\rangle , \\
|\eta \rangle &=&\frac{1}{2}|000\rangle +\frac{1}{2}|101\rangle +\frac{1}{%
\sqrt{2}}|110\rangle , \\
|\xi \rangle &=&\frac{1}{2}|000\rangle +\frac{1}{\sqrt{2}}|101\rangle +\frac{%
1}{2}|110\rangle .
\end{eqnarray*}

It is clear that $|\vartheta ^{\prime }\rangle $, $|\eta ^{\prime }\rangle $%
, and $|\xi ^{\prime }\rangle $ are LU equivalent to their SDs $|\vartheta
\rangle $, $|\eta \rangle $, and $|\xi \rangle $, respectively. It is known
that the 2-tangle $\tau _{xy}$, the negativity, vNEE, and the tangle $\tau
_{x(yz)}$ are LU invariant. Therefore, $|\vartheta ^{\prime }\rangle $ and $%
|\vartheta \rangle $, $|\eta ^{\prime }\rangle $ and $|\eta \rangle $, and $%
|\xi ^{\prime }\rangle $ and $|\xi \rangle $ have the same values for the
four entanglement measures.

We next derive the four entanglement measures via SD. For states of the W
SLOCC\ class, from \cite{Li-qip, Li-qic} the vNEE is

\begin{eqnarray}
S(\rho _{\mu }) &=&-\frac{1+\sqrt{1-4\alpha _{\mu }}}{2}\ln \frac{1+\sqrt{%
1-4\alpha _{\mu }}}{2}  \notag \\
&&-\frac{1-\sqrt{1-4\alpha _{\mu }}}{2}\ln \frac{1-\sqrt{1-4\alpha _{\mu }}}{%
2},  \label{entropy-0}
\end{eqnarray}%
$\mu \in \{A,B,C\}$, $0\leq \alpha _{\mu }\leq 1/4$, where
\begin{equation}
\alpha _{A}=\lambda _{0}^{2}(\lambda _{2}^{2}+\lambda _{3}^{2}),\alpha
_{B}=\lambda _{3}^{2}(\lambda _{0}^{2}+\lambda _{2}^{2}),\alpha _{C}=\lambda
_{2}^{2}(\lambda _{0}^{2}+\lambda _{3}^{2}).  \label{wsc-3}
\end{equation}%
It is known that $S(\rho _{\mu })$ increases strictly monotonically as $%
\alpha _{\mu }$ does \cite{Li-qip}.\ Clearly, the following Proposition 1
holds from \cite{Li-qip, Li-qic}.

\textit{Proposition 1. The }maximal vNEE is $\ln 2$. And $S(\rho _{\mu
})=\ln 2$ if and only if $\alpha _{\mu }=1/4$.

Let $A_{vnee}$ stand for the average vNEE. Then,

\begin{equation}
A_{vnee}=\frac{S(\rho _{A})+S(\rho _{B})+S(\rho _{C})}{3}.  \label{vNEE}
\end{equation}

From \cite{Li-qip}, for states of the W SLOCC\ class, the 2-tangles $\tau
_{xy}$\ are

\begin{equation}
\tau _{AB}=4\lambda _{0}^{2}\lambda _{3}^{2},\tau _{AC}=4\lambda
_{0}^{2}\lambda _{2}^{2},\tau _{BC}=4\lambda _{2}^{2}\lambda _{3}^{2}.
\label{w-3}
\end{equation}%
Let $A\tau _{xy}$ stand for the average 2-tangle. Then,

\begin{equation}
A\tau _{xy}=\frac{\tau _{AB}+\tau _{AC}+\tau _{BC}}{3}.  \label{2-tangle}
\end{equation}

By the definitions \cite{Vidal, Ou}, a calculation yields the following
negativity for the W-like states in Eq. (\ref{w-like}). Let $A_{neg}$ stand
for the average negativity. Then,
\begin{equation}
A_{neg}=\frac{\mathcal{N}_{AB}+\mathcal{N}_{AC}+\mathcal{N}_{BC}}{3},
\label{aneg}
\end{equation}%
where
\begin{eqnarray}
\mathcal{N}_{AB} &=&\frac{1}{2}\left( \sqrt{4\lambda _{3}^{2}\lambda
_{0}^{2}+\lambda _{2}^{4}}-\lambda _{2}^{2})\right) ,  \label{neg-1} \\
\mathcal{N}_{AC} &=&\frac{1}{2}\left( \sqrt{\lambda _{3}^{4}+4\lambda
_{2}^{2}\lambda _{0}^{2}}-\lambda _{3}^{2}\right) ,  \label{neg-2} \\
\mathcal{N}_{BC} &=&\frac{1}{2}\left( \sqrt{4\lambda _{3}^{2}\lambda
_{2}^{2}+\lambda _{0}^{4}}-\lambda _{0}^{2}\right) .  \label{neg-3}
\end{eqnarray}

We derive $\mathcal{N}_{AB}$ below. Let $\rho _{AB}=tr_{C}\allowbreak (|\Psi
\rangle \langle \Psi |)$, where

\begin{equation}
|\Psi \rangle =\lambda _{0}|000\rangle +\lambda _{2}|101\rangle +\lambda
_{3}|110\rangle .
\end{equation}%
A calculation yields the partial transpose $\rho _{AB}^{T_{A}}$\ and the
eigenvalues of $\rho _{AB}^{T_{A}}$ which are $\frac{1}{2}\lambda
_{2}^{2}\pm \frac{1}{2}\sqrt{4\lambda _{3}^{2}\lambda _{0}^{2}+\lambda
_{2}^{4}}$, $\lambda _{0}^{2}$, and $\lambda _{3}^{2}$. By the definition of
the negativity \cite{Vidal}, $\mathcal{N}_{AB}$ is equal to the sum of the
absolute values of the negative eigenvalues of the partial transpose $\rho
_{AB}^{T_{A}}$. Thus, $\mathcal{N}_{AB}=\left( \left\vert \left\vert \rho
_{AB}^{T_{A}}\right\vert \right\vert -1\right) /2=\frac{1}{2}\left( \sqrt{%
4\lambda _{3}^{2}\lambda _{0}^{2}+\lambda _{2}^{4}}-\lambda _{2}^{2})\right)
$. Similarly, we can derive\ $\mathcal{N}_{AC}$ and $\mathcal{N}_{BC}$.

Let $\tau _{x(yz)}$, where $x(yz)=A(BC),B(AC),C(AB)$, stand for the tangle
\cite{Coffman}, and $A\tau _{x(yz)}$ for the average tangle. Then, for any
state of the W SLOCC\ class, a calculation yields

\begin{equation}
A\tau _{x(yz)}=\frac{\tau _{A(BC)}+\tau _{B(AC)}+\tau _{C(AB)}}{3},
\end{equation}%
where

\begin{eqnarray}
\tau _{A(BC)} &=&4\lambda _{0}^{2}(\lambda _{2}^{2}+\lambda _{3}^{2}),
\label{abc-1} \\
\tau _{B(AC)} &=&4\lambda _{3}^{2}(\lambda _{0}^{2}+\lambda _{2}^{2}),
\label{abc-2} \\
\tau _{C(AB)} &=&4\lambda _{2}^{2}\left( \lambda _{0}^{2}+\lambda
_{3}^{2}\right) .  \label{abc-3}
\end{eqnarray}%
Via the above formulas, a straightforward calculation yields the tangle, the
2-tangle, vNEE, the negativity, and their averages for states $|\vartheta
\rangle $, $|\eta \rangle $, $|\xi \rangle $, W, and GHZ in Tables 1-4. Note
that $|\vartheta ^{\prime }\rangle $ and $|\vartheta \rangle $, $|\eta
^{\prime }\rangle $ and $|\eta \rangle $, and $|\xi ^{\prime }\rangle $ and $%
|\xi \rangle $ have the same values in Tables 1-4.

\section{The states having the maximal vNEE $S(\protect\rho _{A})(=\ln 2)$}

\textit{Lemma 1}. For a state of the W SLOCC\ class, its vNEE $S(\rho _{A})$
is $\ln 2$\ if and only if \ under LU\ the state is
\begin{equation}
\frac{1}{\sqrt{2}}|000\rangle +\lambda _{2}|101\rangle +\lambda
_{3}|110\rangle ,  \label{eq-w-8}
\end{equation}%
where $\lambda _{2}^{2}+\lambda _{3}^{2}=1/2$ and $\lambda _{2}\lambda
_{3}\neq 0$.

Proof. By Proposition 1, when $S(\rho _{A})=\ln 2$, then
\begin{equation}
\alpha _{A}=\lambda _{0}^{2}(\lambda _{2}^{2}+\lambda _{3}^{2})=1/4.\text{ }
\label{eq-w-1}
\end{equation}

Then, from $\sum_{i=0}^{3}\lambda _{i}^{2}=1$ in Eq. (\ref{w-0}) and via Eq.
(\ref{eq-w-1}), we obtain

\begin{equation}
\lambda _{0}^{4}-\lambda _{0}^{2}(1-\lambda _{1}^{2})+1/4=0\text{. }
\label{eq-w-3}
\end{equation}

To guarantee $\lambda _{0}\geq 0$, $\lambda _{1}$ in Eq. (\ref{eq-w-3})\
must vanish. Then, from Eq. (\ref{eq-w-3}), we obtain $\lambda _{0}^{2}=1/2$%
. From $\lambda _{0}^{2}+\lambda _{2}^{2}+\lambda _{3}^{2}=1$, we obtain $%
\lambda _{2}^{2}+\lambda _{3}^{2}=1/2$. Thus, Eq. (\ref{w-0}) reduces to Eq.
(\ref{eq-w-8}).

Conversely, for the state in Eq. (\ref{eq-w-8}), a calculation yields $%
S(\rho _{A})=\ln 2$. Q.E.D.

When $\lambda _{2}=\lambda _{3}=1/2$ in Eq. (\ref{eq-w-8}), we obtain the
state $|\vartheta \rangle $. Via Lemma 1, we next investigate the properties
of $|\vartheta \rangle $.

\textit{Property} \textit{1.1.} $|\vartheta \rangle $ is a unique state
which has the maximal average 2-tangle $A\tau _{xy}$ among the states having
$S(\rho _{A})=\ln 2$ (i.e. among the states in Eq. (\ref{eq-w-8})).

Proof. For the states in Eq. (\ref{eq-w-8}),\ a calculation yields that the
average 2-tangle
\begin{equation}
A\tau _{xy}=\frac{1+4\lambda _{2}^{2}\lambda _{3}^{2}}{3},
\end{equation}%
which\ has the maximum $\frac{5}{12}$ at $\lambda _{2}=\lambda _{3}=1/2$.
Thus, Proposition 1.1 is true. Q.E.D.

\textit{Property 1.2.} $|\vartheta \rangle $ is a unique state which has the
maximal average vNEE $A_{vnee}$\ among the states having $S(\rho _{A})=\ln 2$
(i.e. among the states in Eq. (\ref{eq-w-8})).

Proof. Let us compute $\max A_{vnee}$ among the states in Eq. (\ref{eq-w-8}%
). For the states in Eq. (\ref{eq-w-8}), $S(\rho _{A})=\ln 2$, then\ we only
need to compute $\max (S(\rho _{B})+S(\rho _{C}))$. For the states in Eq. (%
\ref{eq-w-8}), $\alpha _{B}=1/4-\lambda _{2}^{4}$ and $\alpha _{C}=\lambda
_{2}^{2}-\lambda _{2}^{4}$. From Eq. (\ref{entropy-0}), $S(\rho _{B})$ and $%
S(\rho _{C})$ are the functions of $\lambda _{2}$. The derivatives of $%
S(\rho _{B})$ and $S(\rho _{C})$\ with respect to $\lambda _{2}$ are

\begin{eqnarray*}
S(\rho _{C})^{\prime } &=&-\frac{2\lambda _{2}-4\lambda _{2}^{3}}{\sqrt{%
1-4(\lambda _{2}^{2}-\lambda _{2}^{4})}}\ln \frac{1-\sqrt{1-4(\lambda
_{2}^{2}-\lambda _{2}^{4})}}{1+\sqrt{1-4(\lambda _{2}^{2}-\lambda _{2}^{4})}}%
, \\
S(\rho _{B})^{\prime } &=&-2\lambda _{2}\ln \frac{1+2\lambda _{2}^{2}}{%
1-2\lambda _{2}^{2}}.
\end{eqnarray*}

When $\lambda _{2}=1/2$, $S(\rho _{B})^{\prime }+S(\rho _{C})^{\prime }=0$.
One can check that $S(\rho _{B})+S(\rho _{C})$ reaches its maximum at $%
\lambda _{2}=1/2$. Thus, $\max A_{vnee}=\frac{10\ln 2-3\ln 3}{6}$ at $%
\lambda _{2}=1/2$. When $\lambda _{2}=1/2$, the state is just $|\vartheta
\rangle $. Thus, Proposition 1.2 is true. Q.E.D.

\textit{Property 1.3.} $|\vartheta \rangle $ is a unique state\ which has
the maximal average negativity $A_{neg}$ among the states having $S(\rho
_{A})=\ln 2$ (i.e. among the states in Eq. (\ref{eq-w-8})).

Proof. For the states in Eq. (\ref{eq-w-8}), a calculation yields that the
average negativity
\begin{eqnarray*}
&&\text{ }A_{neg} \\
&=&\frac{1}{6}(\sqrt{\lambda _{2}^{4}+2\lambda _{3}^{2}}+\sqrt{\lambda
_{3}^{4}+2\lambda _{2}^{2}} \\
&&+\sqrt{4\lambda _{2}^{2}\lambda _{3}^{2}+\frac{1}{4}}-1),
\end{eqnarray*}%
which has the maximum $\frac{\sqrt{2}+1}{12}$ at $\lambda _{3}=\lambda
_{2}=1/2$. Thus, Proposition 1.3 is true. Q.E.D.

\textit{Property 1.4.} $|\vartheta \rangle $ is a unique state\ which has
the maximal average tangle $A\tau _{x(yz)}$ among the states having $S(\rho
_{A})=\ln 2$ (i.e. among the states in Eq. (\ref{eq-w-8})).

Proof. For the states in Eq. (\ref{eq-w-8}), a calculation yields that the
average tangle $A\tau _{x(yz)}$

\begin{equation}
A\tau _{x(yz)}=\frac{8}{3}\left( \frac{1}{2}\lambda _{2}^{2}+\frac{1}{2}%
\lambda _{3}^{2}+\lambda _{2}^{2}\lambda _{3}^{2}\right) ,\text{ }
\end{equation}%
which has the maximum $\frac{5}{6}$ at $\lambda _{3}=\lambda _{2}=1/2$.
Thus, Proposition 1.4 is true. Q.E.D.

From Properties 1.1 to 1.4, we conclude the following.

\textit{Theorem 1}. $|\vartheta \rangle $ is a unique state\ which has the
maximal $A\tau _{xy}$, $A_{vnee}$, $A\tau _{x(yz)}$, and $A_{neg}$ among the
states having $S(\rho _{A})=\ln 2$ (i.e. among the states in Eq. (\ref%
{eq-w-8})).

\section{The states having the maximal vNEE $S(\protect\rho _{B})$ $(=\ln 2)$%
}

\textit{Lemma 2.} For a state of the W SLOCC\ class, its vNEE $S(\rho _{B})$
is $\ln 2$\ if and only if under LU\ the state is

\begin{equation}
\lambda _{0}|000\rangle +\lambda _{2}|101\rangle +\frac{1}{\sqrt{2}}%
|110\rangle ,  \label{w-2-8}
\end{equation}%
where $\lambda _{0}^{2}+\lambda _{2}^{2}=1/2$ and $\lambda _{0}\lambda
_{2}\neq 0$.

Proof. By Proposition 1, when $S(\rho _{B})=\ln 2$, then $\alpha
_{B}=\lambda _{3}^{2}(\lambda _{0}^{2}+\lambda _{2}^{2})=1/4$. From $%
\sum_{i=0}^{3}\lambda _{i}^{2}=1$ in Eq. (\ref{w-0}), we obtain

\begin{equation}
\lambda _{3}^{4}-\lambda _{3}^{2}(1-\lambda _{1}^{2})+1/4=0\text{. }
\label{w-2-3}
\end{equation}

Similarly, to guarantee $\lambda _{3}\geq 0$, $\lambda _{1}$ in Eq. (\ref%
{w-2-3})\ must vanish. Then, from Eq. (\ref{w-2-3}) we obtain $\lambda
_{3}^{2}=1/2$. From $\lambda _{0}^{2}+\lambda _{2}^{2}+\lambda _{3}^{2}=1$,
we obtain $\lambda _{0}^{2}+\lambda _{2}^{2}=1/2$. Thus, Eq. (\ref{w-0})
reduces to Eq. (\ref{w-2-8}).

Conversely, for the state in Eq. (\ref{w-2-8}), a calculation yields $S(\rho
_{B})=\ln 2$. Q.E.D.

When $\lambda _{0}=\lambda _{2}=1/2$ in Eq. (\ref{w-2-8}), we obtain the
state $|\eta \rangle $. Via Lemma 2, similarly, we conclude the following.

\textit{Theorem} \textit{2. }$|\eta \rangle $ is a unique state\ which has
the maximal $A\tau _{xy}$, $A_{vnee}$, $A\tau _{x(yz)}$, and $A_{neg}$\
among the states having $S(\rho _{B})=\ln 2$ (i.e. among the states in Eq. (%
\ref{w-2-8})).

Proof. A complicated calculation yields $\max A\tau _{xy}=\frac{5}{12}$, $%
\max A_{vnee}=\frac{10\ln 2-3\ln 3}{6}$, $\max A\tau _{x(yz)}=\frac{5}{6}$,
and $\max A_{neg}=\frac{\sqrt{2}+1}{12}$ at $\lambda _{0}=\lambda _{2}=1/2$
among the states in Eq. (\ref{w-2-8}). Thus, Theorem 2 holds. Q.E.D.

\section{The states having the maximal vNEE $S(\protect\rho _{C})(=\ln 2)$}

\textit{Lemma 3. }For a state of the W SLOCC\ class, its vNEE $S(\rho _{C})$
is $\ln 2$\ if and only if under LU the state is

\begin{equation}
\lambda _{0}|000\rangle +\frac{1}{\sqrt{2}}|101\rangle +\lambda
_{3}|110\rangle .  \label{w-3-8}
\end{equation}%
where $\lambda _{0}^{2}+\lambda _{3}^{2}=1/2$ and $\lambda _{0}\lambda
_{3}\neq 0$.

Proof. By Proposition 1, when $S(\rho _{C})=\ln 2$, then $\alpha
_{C}=\lambda _{2}^{2}(\lambda _{0}^{2}+\lambda _{3}^{2})=1/4$. Via $%
\sum_{i=0}^{3}\lambda _{i}^{2}=1$ in Eq. (\ref{w-0}), we obtain

\begin{equation}
\lambda _{2}^{4}-\lambda _{2}^{2}(1-\lambda _{1}^{2})+1/4=0\text{. }
\label{w-3-4}
\end{equation}

Similarly, to guarantee $\lambda _{2}\geq 0$, $\lambda _{1}$ in Eq. (\ref%
{w-3-4}) must vanish. Then, from Eq. (\ref{w-3-4}) we obtain $\lambda
_{2}^{2}=1/2$. From $\lambda _{0}^{2}+\lambda _{2}^{2}+\lambda _{3}^{2}=1$,
we obtain $\lambda _{0}^{2}+\lambda _{3}^{2}=1/2$. Thus, Eq. (\ref{w-0})
reduces to Eq. (\ref{w-3-8}).

Conversely, for the state in Eq. (\ref{w-3-8}), a calculation yields $S(\rho
_{C})=\ln 2$. Q.E.D.

When $\lambda _{0}=\lambda _{3}=1/2$ in Eq. (\ref{w-3-8}), we obtain the
state $|\xi \rangle $. Via Lemma 3, similarly, we conclude the following.

\textit{Theorem} \textit{3}.\textit{\ }$|\xi \rangle $ is a unique state\
which has the maximal $A\tau _{xy}$, $A_{vnee}$, $A\tau _{x(yz)}$,\ and $%
A_{neg}$ among the states having $S(\rho _{C})=\ln 2$ (i.e. among the states
in Eq. (\ref{w-3-8})).

Proof. A complicated calculation yields $\max A\tau _{xy}=\frac{5}{12}$, $%
\max A_{vnee}=\frac{10\ln 2-3\ln 3}{6}$, $\max A\tau _{x(yz)}=\frac{5}{6}$,
and $\max A_{neg}=\frac{\sqrt{2}+1}{12}$ at $\lambda _{0}=\lambda _{3}=1/2$
among the states in Eq. (\ref{w-3-8}). Thus, Theorem 3 holds. Q.E.D.

\section{$A_{neg}$, $A\protect\tau _{x(yz)}$, $A_{vnee}$, and $A\protect\tau %
_{xy}$ for the W state}

Note that the W state and its SD\ are LU equivalent.

\textit{Result 1}. $A_{neg}$ for the W state is the maximal among the W-like
states in Eq. (\ref{w-like}).

Proof. For the W-like states in Eq. (\ref{w-like}), by the definition of $%
A_{neg}$,
\begin{equation}
A_{neg}=\left( \frac{\mathcal{N}_{AB}+\mathcal{N}_{AC}+\mathcal{N}_{BC}}{3}%
\right) .
\end{equation}

From Eqs. (\ref{neg-1}-\ref{neg-3}) and via that $\lambda _{3}^{2}+\lambda
_{2}^{2}+\lambda _{0}^{2}=1$, we obtain
\begin{eqnarray*}
&&A_{neg}=\frac{1}{6}(\sqrt{4\lambda _{3}^{2}\lambda _{0}^{2}+\lambda
_{2}^{4}}+\sqrt{\lambda _{3}^{4}+4\lambda _{2}^{2}\lambda _{0}^{2}} \\
&&+\sqrt{4\lambda _{3}^{2}\lambda _{2}^{2}+\lambda _{0}^{4}}-1).
\end{eqnarray*}

A calculation shows that $A_{neg}$ has the conditional extremum $\frac{\sqrt{%
5}-1}{6}$ at $\lambda _{3}=\lambda _{2}=\lambda _{0}=1/\sqrt{3}$, which is
just SD of the W state. One can verify that $\frac{\sqrt{5}-1}{6}$ is the
maximal. That is, $\max A_{neg}=\frac{\sqrt{5}-1}{6}$ at SD of the W state.
Therefore, Result 1 holds. Q.E.D.

\textit{Remark 1.} The average negativity for the W state is not the maximal
even among the W SLOCC\ class.

Let
\begin{equation}
|\varpi \rangle =\frac{\sqrt{21}}{8}(|000\rangle +\frac{1}{\sqrt{21}}%
e^{i\theta }|100\rangle +|101\rangle +|110\rangle ).
\end{equation}

Clearly, $|\varpi \rangle $ belongs to the W SLOCC class. By the definition
of the negativity \cite{Vidal}, a calculation yields the average negativity $%
0.244\,81$ for $|\varpi \rangle $ while $\frac{\sqrt{5}-1}{6}=0.206\,01$ for
the W state.

\textit{Result 2}. $A\tau _{x(yz)}$ for the W state is the maximal among the
W SLOCC class.

Proof. By the definition, $A\tau _{x(yz)}=\frac{\tau _{A(BC)}+\tau
_{B(AC)}+\tau _{C(AB)}}{3}$. For any state of the W SLOCC class in Eq. (\ref%
{w-0}), from Eqs. (\ref{abc-1}-\ref{abc-3}), we obtain
\begin{equation}
A\tau _{x(yz)}=\frac{8}{3}\left( \lambda _{0}^{2}\lambda _{2}^{2}+\lambda
_{0}^{2}\lambda _{3}^{2}+\lambda _{2}^{2}\lambda _{3}^{2}\right) .
\end{equation}

$\allowbreak $Note that $\sum_{i=0}^{3}\lambda _{i}^{2}=1$. A calculation
shows that $A\tau _{x(yz)}$ has the conditional extremum $\frac{8}{9}$\ at $%
\lambda _{3}=\lambda _{2}=\lambda _{0}=1/\sqrt{3}$, which is just SD of the
W state. One can verify that $\frac{8}{9}$ is maximal for $A\tau _{x(yz)}$.
Therefore, Result 2 holds. Q.E.D.

\textit{Result 3}. $A_{vnee}$ for the W state is the maximal among the W
SLOCC class.

Proof. For the W state, $A_{vnee}=\frac{3\ln 3-2\ln 2}{3}$, which is the
maximal among the W SLOCC\ class \cite{Li-qip}. Q.E.D.

Result 4. $A\tau _{xy}$ for the W state is the maximal among all states of
three qubits.

Proof. For the W state, $A\tau _{xy}=4/9$, which is the maximal among all
states of three qubits \cite{Dur}. Q.E.D.

\section{comparisons and explanations}

Table 1. The 2-tangle $\tau _{xy}$ and $A\tau _{xy}$

\begin{tabular}{|l|l|l|l|l|}
\hline
& $\tau _{AB}$ & $\tau _{AC}$ & $\tau _{BC}$ & $A\tau _{xy}$ \\ \hline
GHZ & 0 & 0 & 0 & 0 \\ \hline
W & $\frac{4}{9}$ & $\frac{4}{9}$ & $\frac{4}{9}$ & $\frac{4}{9}$ (max) \\
\hline
$|\vartheta \rangle ,|\vartheta ^{\prime }\rangle $ & $\frac{1}{2}$ & $\frac{%
1}{2}$ & $\frac{1}{4}$ & $\frac{5}{12}$ \\ \hline
$|\eta \rangle ,|\eta ^{\prime }\rangle $ & $\frac{1}{2}$ & $\frac{1}{4}$ & $%
\frac{1}{2}$ & $\frac{5}{12}$ \\ \hline
$|\xi \rangle ,|\xi ^{\prime }\rangle $ & $\frac{1}{4}$ & $\frac{1}{2}$ & $%
\frac{1}{2}$ & $\frac{5}{12}$ \\ \hline
\end{tabular}

Note that $\frac{1}{2}>\frac{4}{9}$, $\frac{4}{9}=\frac{16}{36}$, and $\frac{%
5}{12}=\frac{15}{36}$.

Comparisons and explanations via Table 1 and 2-tangle:

(1.1). The first two rows say when one qubit is traced out, then the state
of the remaining 2-qubit system is still entangled for the W state while
separable for the GHZ state.

(1.2). The last column says that the states W, $|\vartheta ^{\prime }\rangle
$, $|\vartheta \rangle $, $|\eta ^{\prime }\rangle $, $|\eta \rangle $, $%
|\xi ^{\prime }\rangle $, and $|\xi \rangle $ have almost the same average
2-tangle.

(1.3). The third row says that for the states $|\vartheta ^{\prime }\rangle $
and $|\vartheta \rangle $,\ when any one of qubits $C$ and $B$\ is traced
out, then the remaining two qubits are more entangled than the two qubits
for the W state.

(1.4). The fourth row says that for the states $|\eta ^{\prime }\rangle $
and $|\eta \rangle $, when any one of qubits $A$ and $C$\ is traced out,
then the remaining two qubits are more entangled than the two qubits for the
W state.

(1.5). The fifth row says that for the states $|\xi ^{\prime }\rangle $, and
$|\xi \rangle $, when any one of qubits $A$ and $B$ is traced out, then the
remaining two qubits are more entangled than the two qubits for the W state.

Table 2. vNEE and $A_{vnee}$

\begin{tabular}{|l|l|l|l|l|}
\hline
& $S(\rho _{A})$ & $S(\rho _{B})$ & $S(\rho _{C})$ & $A_{vnee}$ \\ \hline
GHZ & $\ln 2$ (max) & $\ln 2$ & $\ln 2$ & $\ln 2$ \\ \hline
W & $\frac{3\ln 3-2\ln 2}{3}$ & $\frac{3\ln 3-2\ln 2}{3}$ & $\frac{3\ln
3-2\ln 2}{3}$ & $\frac{3\ln 3-2\ln 2}{3}$ \\ \hline
$|\vartheta \rangle ,|\vartheta ^{\prime }\rangle $ & $\ln 2$ (max) & $\frac{%
8\ln 2-3\ln 3}{4}$ & $\frac{8\ln 2-3\ln 3}{4}$ & $\frac{10\ln 2-3\ln 3}{6}$
\\ \hline
$|\eta \rangle ,|\eta ^{\prime }\rangle $ & $\frac{8\ln 2-3\ln 3}{4}$ & $\ln
2$ (max) & $\frac{8\ln 2-3\ln 3}{4}$ & $\frac{10\ln 2-3\ln 3}{6}$ \\ \hline
$|\xi \rangle ,|\xi ^{\prime }\rangle $ & $\frac{8\ln 2-3\ln 3}{4}$ & $\frac{%
8\ln 2-3\ln 3}{4}$ & $\ln 2$ (max) & $\frac{10\ln 2-3\ln 3}{6}$ \\ \hline
\end{tabular}

Note that $\ln 2=\allowbreak 0.693,$ $\frac{3\ln 3-2\ln 2}{3}=\allowbreak
0.637,$ $\frac{8\ln 2-3\ln 3}{4}=\allowbreak 0.562$, $\frac{10\ln 2-3\ln 3}{6%
}=\allowbreak 0.606$.

Comparisons and explanations via Table 2 and vNEE:

(2.1). The first two rows say that the single-qubit state$\ \rho _{x}$,
where $x=A$, $B$, and $C$, is the maximally mixed state for the GHZ state
but not for the W state.

(2.2). The last column says that W, $|\vartheta ^{\prime }\rangle $, $%
|\vartheta \rangle $, $|\eta ^{\prime }\rangle $, $|\eta \rangle $, $|\xi
^{\prime }\rangle $, and $|\xi \rangle $ have almost the same average von
Neumann entanglement entropy.

(2.3). The third row says that the single-qubit state$\ \rho _{A}$ is the
maximally mixed for $|\vartheta ^{\prime }\rangle $ and $|\vartheta \rangle $
but not for W state.

(2.4). The fourth row says that the single-qubit state$\ \rho _{B}$ is the
maximally mixed for $|\eta ^{\prime }\rangle $ and $|\eta \rangle $ but not
for W state.

(2.5). The fifth row says that the single-qubit state$\ \rho _{C}$ is the
maximally mixed for $|\xi ^{\prime }\rangle $ and $|\xi \rangle $ but not
for W state.

Table 3. For the negativity and $A_{neg}$

\begin{tabular}{|l|l|l|l|l|}
\hline
& $\mathcal{N}_{AB}$ & $\mathcal{N}_{AC}$ & $\mathcal{N}_{BC}$ & $A_{neg}$
\\ \hline
GHZ & 0 & 0 & 0 & 0 \\ \hline
W & $\frac{\sqrt{5}-1}{6}$ & $\frac{\sqrt{5}-1}{6}$ & $\frac{\sqrt{5}-1}{6}$
& $\frac{\sqrt{5}-1}{6}$ \\ \hline
$|\vartheta \rangle ,|\vartheta ^{\prime }\rangle $ & $\frac{1}{4}$ & $\frac{%
1}{4}$ & $\frac{\sqrt{2}-1}{4}$ & $\frac{\sqrt{2}+1}{12}$ \\ \hline
$|\eta \rangle ,|\eta ^{\prime }\rangle $ & $\frac{1}{4}$ & $\frac{\sqrt{2}-1%
}{4}$ & $\frac{1}{4}$ & $\frac{\sqrt{2}+1}{12}$ \\ \hline
$|\xi \rangle ,|\xi ^{\prime }\rangle $ & $\frac{\sqrt{2}-1}{4}$ & $\frac{1}{%
4}$ & $\frac{1}{4}$ & $\frac{\sqrt{2}+1}{12}$ \\ \hline
\end{tabular}%
{}

Here, $\frac{\sqrt{5}-1}{6}=\allowbreak 0.206<\frac{1}{4},\,\frac{\sqrt{2}-1%
}{4}=\allowbreak 0.104$, $\frac{\sqrt{2}+1}{12}=\allowbreak 0.201$.

Comparisons and explanations via Table 3 and the negativity:

For W-like states, from Eqs. (\ref{w-3}, \ref{neg-1}-\ref{neg-3}), we obtain
\begin{eqnarray}
\tau _{AB} &=&4\mathcal{N}_{AB}^{2}+4\lambda _{2}^{2}\mathcal{N}_{AB}, \\
\tau _{AC} &=&4\mathcal{N}_{AC}^{2}+4\lambda _{3}^{2}\mathcal{N}_{AC}, \\
\tau _{BC} &=&4\mathcal{N}_{BC}^{2}+4\lambda _{0}^{2}\mathcal{N}_{BC}.
\end{eqnarray}

Thus, we obtain

(3.1) is the same as (1.1).

(3.2). The last column says that the states W, $|\vartheta ^{\prime }\rangle
$, $|\vartheta \rangle $, $|\eta ^{\prime }\rangle $, $|\eta \rangle $, $%
|\xi ^{\prime }\rangle $, and $|\xi \rangle $ have almost the same average
negativity.

(3.3), (3.4), and (3.5) are the same as (1.3), (1.4), and (1.5),
respectively.

Table 4. For tangles $\tau _{x(yz)}$ and $A\tau _{x(yz)}$

\begin{tabular}{|l|l|l|l|l|}
\hline
& $\tau _{A(BC)}$ & $\tau _{B(AC)}$ & $\tau _{C(AB)}$ & $A\tau _{x(yz)}$ \\
\hline
GHZ & 1 & 1 & 1 & 1 \\ \hline
W & $\frac{8}{9}$ & $\frac{8}{9}$ & $\frac{8}{9}$ & $\frac{8}{9}$ \\ \hline
$|\vartheta \rangle ,|\vartheta ^{\prime }\rangle $ & 1(max) & $\frac{3}{4}$
& $\frac{3}{4}$ & $\frac{5}{6}$ \\ \hline
$|\eta \rangle ,|\eta ^{\prime }\rangle $ & $\frac{3}{4}$ & 1(max) & $\frac{3%
}{4}$ & $\frac{5}{6}$ \\ \hline
$|\xi \rangle ,|\xi ^{\prime }\rangle $ & $\frac{3}{4}$ & $\frac{3}{4}$ &
1(max) & $\frac{5}{6}$ \\ \hline
\end{tabular}

Comparisons and explanations via Table 4 and the tangles $\tau _{x(yz)}$:

(4.1). The first two rows say that the entanglement between qubit $x$\ \ and
qubits $y$\ and $z$\ is maximal for the GHZ state but not for the W state.

(4.2). The last column says that W, $|\vartheta ^{\prime }\rangle $, $%
|\vartheta \rangle $, $|\eta ^{\prime }\rangle $, $|\eta \rangle $, $|\xi
^{\prime }\rangle $, and $|\xi \rangle $ have almost the same average tangle
$\tau _{x(yz)}$.

(4.3). The third row says that the entanglement between\ qubit $A$ and
qubits $B$\ and $C$\ is maximal for $|\vartheta ^{\prime }\rangle $ and $%
|\vartheta \rangle $ but not for W state.

(4.4). The fourth row says that the entanglement between\ qubit $B$ and
qubits $A$\ and $C$\ is maximal for $|\eta ^{\prime }\rangle $ and $|\eta
\rangle $ but not for W state.

(4.5). The fifth row says that the entanglement between qubit $C$ and qubits
$A$\ and $B$\ is maximal for $|\xi ^{\prime }\rangle $ and $|\xi \rangle $
but not for W state.

Let $E_{MW}$ denote the global entanglement measure proposed by Meyer and
Wallach \cite{Meyer} and $E_{AD}$ be the average determinant of the reduced
density matrices for each qubit as a global entanglement measure in \cite%
{Dlipla}. A calculation yields $E_{MW}=E_{AD}=8/9$ for the W state while $%
E_{MW}=E_{AD}=5/8$ for $|\vartheta ^{\prime }\rangle $, $|\eta ^{\prime
}\rangle $, $|\xi ^{\prime }\rangle $, $|\vartheta \rangle $, $|\eta \rangle
$, $|\xi \rangle $.

\section{$|\protect\vartheta \rangle $, $|\protect\eta \rangle $, $|\protect%
\xi \rangle $, $|\protect\vartheta ^{\prime }\rangle $, $|\protect\eta %
^{\prime }\rangle $, and $|\protect\xi ^{\prime }\rangle $\ can be used for
perfect teleportation.}

Alice has a qubit `a' in the unknown state $|\varphi \rangle _{a}=(\alpha
|0\rangle _{a}+\beta |1\rangle _{a})$ and wants to send the unknown state $%
|\varphi \rangle _{a}$ to Bob. In \cite{Pankaj, LLI, Dli-aop}, the authors
indicated that the W state can not be suitable for perfect teleportation.
Via the following theorem 4, we can show that Alice can send the unknown
state $|\varphi \rangle _{a}$ to Bob by sharing the states $|\vartheta
\rangle $, $|\eta \rangle $, $|\xi \rangle $, $|\vartheta ^{\prime }\rangle $%
, $|\eta ^{\prime }\rangle $, and $|\xi ^{\prime }\rangle $, respectively.

\textit{Theorem 4}. $|\vartheta \rangle $, $|\eta \rangle $, $|\xi \rangle $%
, $|\vartheta ^{\prime }\rangle $, $|\eta ^{\prime }\rangle $, and $|\xi
^{\prime }\rangle $\ are suitable for perfect teleportation.

Proof. We show that $|\xi \rangle _{123}$ and $|\xi ^{\prime }\rangle _{123}$
are suitable for perfect teleportation below, and that $|\vartheta \rangle
_{123}$ and $|\vartheta ^{\prime }\rangle _{123}$, and $|\eta \rangle $\ and
$|\eta ^{\prime }\rangle $\ are suitable for perfect teleportation in
Appendix A.

In \cite{Pankaj}, the authors demonstrated how to teleport the unknown state
$|\varphi \rangle _{a}$\ by sharing $|\xi ^{\prime }\rangle _{123}$, where
Alice has qubits `1' and `2' while Bob has the qubit `3'.

Next, let Alice and Bob share the state $|\xi \rangle _{123}$, where Alice
has qubits `1' and `2' while Bob has the qubit `3'. Alice also has qubit
`a'. By Result 2 of \cite{Dli-aop}, $|\xi \rangle _{123}$ is a suitable
resource to teleport the unknown state $|\varphi \rangle _{a}$. For
readability, we give a detailed derivation to show that $|\xi \rangle _{123}$
can be used for the teleportation below.

We can rewrite $|\varphi \rangle _{a}|\xi \rangle _{123}$ as follows.

\begin{eqnarray*}
&&|\varphi \rangle _{a}|\xi \rangle _{123} \\
&=&\frac{1}{2}|000\rangle _{a12}(\alpha |0\rangle _{3})+\frac{1}{2}%
|011\rangle _{a12}(\alpha |0\rangle _{3}) \\
&&+\frac{1}{\sqrt{2}}|110\rangle _{a12}(\beta |1\rangle _{3})+\frac{1}{2}%
|100\rangle _{a12}(\beta |0\rangle _{3}) \\
&&+\frac{1}{2}|111\rangle _{a12}(\beta |0\rangle _{3})+\frac{1}{\sqrt{2}}%
|010\rangle _{a12}(\alpha |1\rangle _{3}) \\
&=&\frac{1}{2}[|\varpi ^{+}\rangle _{a12}(\alpha |0\rangle _{3}+\beta
|1\rangle _{3}) \\
&&+|\varpi ^{-}\rangle _{a12}(\alpha |0\rangle _{3}-\beta |1\rangle _{3}) \\
&&+|\varkappa ^{+}\rangle _{a12}(\beta |0\rangle _{3}+\alpha |1\rangle _{3})
\\
&&+|\varkappa ^{-}\rangle _{a12}(\beta |0\rangle _{3}-\alpha |1\rangle
_{3})],
\end{eqnarray*}%
where
\begin{eqnarray*}
|\varpi ^{\pm }\rangle _{a12} &=&\frac{1}{2}|000\rangle _{a12}+\frac{1}{2}%
|011\rangle _{a12}+\pm \frac{1}{\sqrt{2}}|110\rangle _{a12}, \\
|\varkappa ^{\pm }\rangle _{a12} &=&\pm \frac{1}{\sqrt{2}}|010\rangle _{a12}+%
\frac{1}{2}|100\rangle _{a12}+\frac{1}{2}|111\rangle _{a12}.
\end{eqnarray*}

It is easy to see that $|\varpi ^{\pm }\rangle _{a12}$ and $|\varkappa ^{\pm
}\rangle _{a12}$ are mutually orthogonal. Note that $\sigma _{3}(\alpha
|0\rangle _{3}-\beta |1\rangle _{3})=$ $\sigma _{1}(\beta |0\rangle
_{3}+\alpha |1\rangle _{3})=$ $\sigma _{1}\sigma _{3}(\beta |0\rangle
_{3}-\alpha |1\rangle _{3})=\alpha |0\rangle _{3}+\beta |1\rangle _{3}$. \
Alice can make a von Neumann measurement on the three qubits `a12' by using
the orthogonal states $|\varpi ^{\pm }\rangle _{a12}$ and $|\varkappa ^{\pm
}\rangle _{a12}$ \ and then convey the measurement results to Bob via 2-bit
classical communication \{00 to $|\varpi ^{+}\rangle _{a12}$; 01 to $|\varpi
^{-}\rangle _{a12}$; 10 to $|\varkappa ^{+}\rangle _{a12}$, 11 to $%
|\varkappa ^{-}\rangle _{a12}$\}. Then, via \{00 to $I;$ 01 to $\sigma _{3}$%
; 10 to $\sigma _{1}$; 11 to $\sigma _{1}\sigma _{3}$\} Bob can apply one of
the local unitary operations \{$\mathit{I,}$ $\sigma _{3},\sigma _{1},\sigma
_{1}\sigma _{3}$\} to his qubit, then he converts the state of his qubit `3'
to that of the qubit `a'. Q.E.D.

\section{$|\protect\vartheta \rangle $, $|\protect\eta \rangle $, $|\protect%
\xi \rangle $, $|\protect\vartheta ^{\prime }\rangle $, $|\protect\eta %
^{\prime }\rangle $, and $|\protect\xi ^{\prime }\rangle $ can be used for
perfect superdense coding.}

In \cite{Pankaj}, the authors indicated that the W state can not be suitable
for perfect superdense coding. The following theorem 5 demonstrates how the
protocol works when $|\vartheta \rangle $, $|\eta \rangle $, $|\xi \rangle $%
, $|\vartheta ^{\prime }\rangle $, $|\eta ^{\prime }\rangle $, and $|\xi
^{\prime }\rangle $ are used as resource states for perfect superdense
coding, respectively.

\textit{Theorem 5}. $|\vartheta \rangle $, $|\eta \rangle $, $|\xi \rangle $%
, $|\vartheta ^{\prime }\rangle $, $|\eta ^{\prime }\rangle $, and $|\xi
^{\prime }\rangle $\ are suitable for perfect superdense coding.

Proof. We show that $|\vartheta \rangle _{123}$ and $|\vartheta ^{\prime
}\rangle _{123}$ are suitable for perfect superdense coding below, and that $%
|\eta \rangle $\ and $|\eta ^{\prime }\rangle $ and $|\xi \rangle _{123}$
and $|\xi ^{\prime }\rangle _{123}$\ are suitable for perfect superdense
coding in Appendix B.

In \cite{Pankaj}, the authors demonstrated how their protocol works when $%
|\vartheta ^{\prime }\rangle $ is used as a resource state for perfect
superdense coding, where Alice has qubit 1 while Bob has qubits 2 and 3.

Let Alice and Bob share $|\vartheta \rangle _{123}$, where Alice has qubit 1
while Bob has qubits 2 and 3. By Result 7 of \cite{Dli-aop}, $|\vartheta
\rangle _{123}$ is a suitable resource state for perfect superdense coding.
For readability, we give a detailed derivation to show that $|\vartheta
\rangle _{123}$ can be used for perfect superdense coding below.

Alice can apply unitary operators in \{$I,\sigma _{3},\sigma _{1},\sigma
_{3}\sigma _{1}$\} on her qubit. That is, Alice can apply the following
local unitary operators $A_{i}^{(1)}$, $i=1,2,3,4$, \ to the state $%
|\vartheta \rangle $.

\begin{eqnarray}
A_{1}^{(1)} &=&I\otimes I\otimes I, \\
A_{2}^{(1)} &=&\sigma _{3}\otimes I\otimes I, \\
A_{3}^{(1)} &=&\sigma _{1}\otimes I\otimes I, \\
A_{4}^{(1)} &=&\sigma _{3}\sigma _{1}\otimes I\otimes I.
\end{eqnarray}

Thus, we obtain the following states $A_{i}^{(1)}|\vartheta \rangle $, $%
i=1,2,3,4$,
\begin{eqnarray}
A_{1}^{(1)}|\vartheta \rangle &=&\frac{1}{\sqrt{2}}|000\rangle +\frac{1}{2}%
|101\rangle +\frac{1}{2}|110\rangle , \\
A_{2}^{(1)}|\vartheta \rangle &=&\frac{1}{\sqrt{2}}|000\rangle -\frac{1}{2}%
|101\rangle -\frac{1}{2}|110\rangle , \\
A_{3}^{(1)}|\vartheta \rangle &=&\frac{1}{\sqrt{2}}|100\rangle +\frac{1}{2}%
|001\rangle +\frac{1}{2}|010\rangle , \\
A_{4}^{(1)}|\vartheta \rangle &=&-\frac{1}{\sqrt{2}}|100\rangle +\frac{1}{2}%
|001\rangle +\frac{1}{2}|010\rangle .
\end{eqnarray}

It is easy to check that $A_{i}^{(1)}|\vartheta \rangle $, $i=1,2,3,4$, are
mutually orthogonal. After Alice sends her qubit to Bob, Bob can make a
three-particle von-Neumann measurement on qubits 1, 2 and 3 using the four
mutually orthogonal states $A_{i}^{(1)}|\vartheta \rangle $, $i=1,2,3,4$.
Thus, Bob can determine what unitary operator Alice has applied. It means
that $|\vartheta \rangle $ can be used to transmit 2-bit classical bits by
sending one qubit. Q.E.D.

\section{Summary}

We construct three W-like states $|\vartheta ^{\prime }\rangle $, $|\eta
^{\prime }\rangle $ and $|\xi ^{\prime }\rangle $. $|\vartheta \rangle $
(resp. $|\eta \rangle $ and $|\xi \rangle $) is SD of $|\vartheta ^{\prime
}\rangle $ (resp. $|\eta ^{\prime }\rangle $ and $|\xi ^{\prime }\rangle $).
Thus, $|\vartheta ^{\prime }\rangle $ and $|\vartheta \rangle $, $|\eta
^{\prime }\rangle $ and $|\eta \rangle $, and $|\xi ^{\prime }\rangle $ and $%
|\xi \rangle $ have the same values for the four entanglement measures
because these pairs of states are LU equivalent. Clearly, it is simpler to
compute the four entanglement measures for $|\vartheta \rangle $, $|\eta
\rangle $ and $|\xi \rangle $. We also call $|\vartheta \rangle $, $|\eta
\rangle $ and $|\xi \rangle $ W-like states.

We explore the entanglement properties of $|\vartheta ^{\prime }\rangle $, $%
|\eta ^{\prime }\rangle $and $|\xi ^{\prime }\rangle $ via four entanglement
measures. We show that $|\vartheta \rangle $ (resp. $|\eta \rangle $ and $%
|\xi \rangle $)\ is a unique state which has the maximal average entanglement%
$\ \ A\tau _{xy}$, $A_{vnee}$, $A\tau _{x(yz)}$,\ and $A_{neg}$ among the
states in Eq. (\ref{eq-w-8}) (resp. Eq. (\ref{w-2-8}) and Eq. (\ref{w-3-8}))
by Theorems 1 (resp. 2 and 3) and indicate that $|\vartheta \rangle $, $%
|\eta \rangle $, $|\xi \rangle $, and W have almost the same average
entanglement$\ A\tau _{xy}$, $A_{vnee}$, $A\tau _{x(yz)}$, and $A_{neg}$.

We compare $|\vartheta \rangle $, $|\eta \rangle $, and $|\xi \rangle $ with
the W state via the four measures.

(i). From Table 2, one can know that for the state $|\vartheta \rangle $
(resp. $|\eta \rangle $ and $|\xi \rangle $), the single-qubit state$\ \rho
_{A}$ (resp. $\rho _{B}$ and $\rho _{C}$) is the maximally mixed but not for
the W state via vNEE.

(ii). From Table 4, via the measure $\tau _{x(yz)}$ one can see that for the
state $|\vartheta \rangle $ (resp. $|\eta \rangle $ and $|\xi \rangle $),
the entanglement between\ qubit $A$ (resp. $B$, and $C$) and qubits $B$\ and
$C$\ (resp. $A$ and $C$, and $A$\ and $B$) is maximal but not for W state.

(iii). It is known that tracing out any one of three qubits, the remaining
state is completely unentangled for the GHZ state while maintains
entanglement for the W state \cite{Dur}. Furthermore, from Tables 1 and 3,
one can observe when any one of qubits $C$ and $B$ (resp. $C$ and $A$,\ and $%
A$ and $B$)\ is traced out, then the remaining two qubits for $|\vartheta
\rangle $ (resp. $|\eta \rangle $ and $|\xi \rangle $) are more entangled
than the two qubits for the W state via the 2-tangle and negativity.

For example, for the 2-tangle, $\tau _{AB}=\tau _{AC}=\frac{1}{2}$ for the
state $|\vartheta \rangle $ while $\tau _{AB}=\tau _{AC}=\frac{4}{9}$ for
the W state. For the negativity, $\mathcal{N}_{AB}=\mathcal{N}_{AC}=\frac{1}{%
4}$ for the state $|\vartheta \rangle $ while $\mathcal{N}_{AB}=\mathcal{N}%
_{AC}=\frac{\sqrt{5}-1}{6}$ for the W state. Thus, via the 2-tangle and
negativity, qubits A and B (A and C) of $|\vartheta \rangle $ are more
entangled than qubits A and B (A and C) of the W state.

From the above, one can see that for $|\vartheta \rangle $, $|\eta \rangle $
and $|\xi \rangle $,\ their advantage is higher robustness against some
particle loss than for the W state. It makes $|\vartheta \rangle $, $|\eta
\rangle $ and $|\xi \rangle $ more suitable for transmitting information
over noisy quantum channels, reliable quantum communication networks, and
distributed quantum metrology. Due to the robustness against some particle
loss and the maximal mixedness of the single-qubit state, the three W-like
states may offer stronger advantages in quantum information processing tasks
than the W state. For example, we show that $|\vartheta \rangle $, $|\eta
\rangle $, $|\xi \rangle $, $|\vartheta ^{\prime }\rangle $, $|\eta ^{\prime
}\rangle $, and $|\xi ^{\prime }\rangle $ can be suitable for perfect
teleportation and superdense coding but the W state cannot.

\section{Appendix A. For perfect teleportation}

\setcounter{equation}{0} \renewcommand{\theequation}{A\arabic{equation}}

Alice has a qubit `a' in the unknown state $|\varphi \rangle _{a}=(\alpha
|0\rangle _{a}+\beta |1\rangle _{a})$ and wants to send the unknown state $%
|\varphi \rangle _{a}$ to Bob by sharing the states $|\vartheta \rangle $, $%
|\eta \rangle $, $|\vartheta ^{\prime }\rangle $, and $|\eta ^{\prime
}\rangle $, respectively.

\subsection{Teleport the unknown state $|\protect\varphi \rangle _{a}$\ by
sharing $|\protect\vartheta \rangle _{123}$ and $|\protect\vartheta ^{\prime
}\rangle _{123}$, respectively}

Let Alice and Bob share the state $|\vartheta \rangle _{123}$, where Alice
has qubits `2' and `3' while Bob has the qubit `1'. Alice also has qubit
`a'. We can rewrite $|\varphi \rangle _{a}|\vartheta \rangle _{123}$ as
follows.

\begin{eqnarray*}
&&|\varphi \rangle _{a}|\vartheta \rangle _{123} \\
&=&\frac{1}{\sqrt{2}}|000\rangle _{a23}(\alpha |0\rangle _{1})+\frac{1}{2}%
|001\rangle _{a23}(\alpha |1\rangle _{1}) \\
&&+\frac{1}{2}|010\rangle _{a23}(\alpha |1\rangle _{1})+\frac{1}{\sqrt{2}}%
|100\rangle _{a23}(\beta |0\rangle _{1}) \\
&&+\frac{1}{2}|101\rangle _{a23}(\beta |1\rangle _{1})+\frac{1}{2}%
|110\rangle _{a23}(\beta |1\rangle _{1}) \\
&=&\frac{1}{2}[|\varsigma ^{+}\rangle _{a23}(\alpha |0\rangle _{1}+\beta
|1\rangle _{1}) \\
&&+|\varsigma ^{-}\rangle _{a23}(\alpha |0\rangle _{1}-\beta |1\rangle _{1})
\\
&&+|\upsilon ^{+}\rangle _{a23}(\beta |0\rangle _{1}+\alpha |1\rangle _{1})
\\
&&+|\upsilon ^{-}\rangle _{a23}(\beta |0\rangle _{1}-\alpha |1\rangle _{1})]
\\
&=&\frac{1}{2}[|\varsigma ^{+}\rangle _{a23}(I|\varphi \rangle
_{1})+|\varsigma ^{-}\rangle _{a23}(\sigma _{3}|\varphi \rangle _{1}) \\
&&+|\upsilon ^{+}\rangle _{a23}(\sigma _{1}|\varphi \rangle _{1})+|\upsilon
^{-}\rangle _{a23}(\sigma _{3}\sigma _{1}|\varphi \rangle _{1})],
\end{eqnarray*}%
where

\begin{eqnarray*}
|\varsigma ^{\pm }\rangle _{a23} &=&\frac{1}{\sqrt{2}}|000\rangle _{a23}\pm
\frac{1}{2}|101\rangle _{a23}\pm \frac{1}{2}|110\rangle _{a23}, \\
|\upsilon ^{\pm }\rangle _{a23} &=&\pm \frac{1}{2}|001\rangle _{a23}\pm
\frac{1}{2}|010\rangle _{a23}+\frac{1}{\sqrt{2}}|100\rangle _{a23}.
\end{eqnarray*}

It is easy to see that $|\varsigma ^{\pm }\rangle _{a23}$ and $|\upsilon
^{\pm }\rangle _{a23}$ are mutually orthogonal. Alice can make a von Neumann
measurement on the three particles `a23' by using the orthogonal states $%
|\varsigma ^{\pm }\rangle _{a23}$ and $|\upsilon ^{\pm }\rangle _{a23}$\ and
then convey the measurement results to Bob via 2-bit classical communication
\{00 to $|\varsigma ^{+}\rangle _{a12}$; 01 to $|\varsigma ^{-}\rangle
_{a12} $; 10 to $|\upsilon ^{+}\rangle _{a12}$, 11 to $|\upsilon ^{-}\rangle
_{a12}$\}. Then, via \{00 to $I;$ 01 to $\sigma _{3}$; 10 to $\sigma _{1}$;
11 to $\sigma _{1}\sigma _{3}$\} Bob can apply one of the local unitary
operations \{$\mathit{I,}$ $\sigma _{3},\sigma _{1},\sigma _{1}\sigma _{3}$%
\} to his qubit, then he converts the state of his qubit `1' to that of the
qubit `a'.

Similarly, a calculation yields

\begin{eqnarray*}
&&|\varphi \rangle _{a}|\vartheta ^{\prime }\rangle _{123} \\
&=&(\alpha |0\rangle _{a}+\beta |1\rangle _{a})|\vartheta ^{\prime }\rangle
_{123} \\
&=&\frac{1}{2}[|\varsigma ^{\prime +}\rangle _{a13}(\alpha |0\rangle
_{1}+\beta |1\rangle _{1}) \\
&&+|\varsigma ^{\prime -}\rangle _{a13}(\alpha |0\rangle _{1}-\beta
|1\rangle _{1}) \\
&&+|\upsilon ^{\prime +}\rangle _{a23}(\beta |0\rangle _{1}+\alpha |1\rangle
_{1}) \\
&&+|\upsilon ^{\prime -}\rangle _{a23}(\beta |0\rangle _{1}-\alpha |1\rangle
_{1})],
\end{eqnarray*}%
where
\begin{eqnarray*}
|\varsigma ^{\prime \pm }\rangle _{a23} &=&\frac{1}{2}|001\rangle _{a23}+%
\frac{1}{2}|010\rangle _{a23}\pm \frac{1}{\sqrt{2}}|100\rangle _{a23} \\
|\upsilon ^{\prime \pm }\rangle _{a23} &=&\pm \frac{1}{\sqrt{2}}|000\rangle
_{a23}+\frac{1}{2}|101\rangle _{a23}+\frac{1}{2}|110\rangle _{a23}
\end{eqnarray*}

Via the above discussion for $|\vartheta \rangle _{123}$, Alice can teleport
the unknown state $|\varphi \rangle _{a}$\ by sharing $|\vartheta ^{\prime
}\rangle _{123}$, where Alice has qubits `2' and `3' while Bob has the qubit
`1'.

\subsection{Teleport the unknown state $|\protect\varphi \rangle _{a}$\ by
sharing $|\protect\eta \rangle _{123}$ and $|\protect\eta ^{\prime }\rangle
_{123}$, respectively.}

Let Alice and Bob share the state $|\eta \rangle _{123}$, where Alice has
qubits `1' and `3' while Bob has the qubit `2'. Alice also has qubit `a'. We
can rewrite $|\varphi \rangle _{a}|\eta \rangle _{123}$ as follows.

\begin{eqnarray*}
&&|\varphi \rangle _{a}|\eta \rangle _{123} \\
&=&\frac{1}{2}|000\rangle _{a13}(\alpha |0\rangle _{2})+\frac{1}{2}%
|011\rangle _{a13}(\alpha |0\rangle _{2}) \\
&&+\frac{1}{\sqrt{2}}|010\rangle _{a13}(\alpha |1\rangle _{2})+\frac{1}{2}%
|100\rangle _{a13}(\beta |0\rangle _{2}) \\
&&+\frac{1}{2}|111\rangle _{a13}(\beta |0\rangle _{2})+\frac{1}{\sqrt{2}}%
|110\rangle _{a13}(\beta |1\rangle _{2}) \\
&=&\frac{1}{2}[|\pi ^{+}\rangle _{a13}(\alpha |0\rangle _{2}+\beta |1\rangle
_{2}) \\
&&+|\pi ^{-}\rangle _{a13}(\alpha |0\rangle _{2}-\beta |1\rangle _{2}) \\
&&+|\mu ^{+}\rangle _{a13}(\beta |0\rangle _{2}+\alpha |1\rangle _{2}) \\
&&+|\mu ^{-}\rangle _{a13}(\beta |0\rangle _{2}-\alpha |1\rangle _{2})] \\
&=&\frac{1}{2}[|\pi ^{+}\rangle _{a13}(I|\varphi \rangle _{2})+|\pi
^{-}\rangle _{a13}(\sigma _{3}|\varphi \rangle _{2}) \\
&&+|\mu ^{+}\rangle _{a13}(\sigma _{1}|\varphi \rangle _{2})+|\mu
^{-}\rangle _{a13}(\sigma _{3}\sigma _{1})|\varphi \rangle _{2})].
\end{eqnarray*}%
where

\begin{eqnarray*}
|\pi ^{\pm }\rangle _{a13} &=&\frac{1}{2}|000\rangle _{a13}+\frac{1}{2}%
|011\rangle _{a13}\pm \frac{1}{\sqrt{2}}|110\rangle _{a13} \\
|\mu ^{\pm }\rangle _{a13} &=&\pm \frac{1}{\sqrt{2}}|010\rangle _{a13}+\frac{%
1}{2}|100\rangle _{a13}+\frac{1}{2}|111\rangle _{a13}
\end{eqnarray*}

It is easy to see that $|\pi ^{\pm }\rangle _{a13}$ and $|\mu ^{\pm }\rangle
_{a13}$ are mutually orthogonal. Alice can make a von Neumann measurement on
the three particles `a13' by using the orthogonal states $|\pi ^{\pm
}\rangle _{a13}$ and $|\mu ^{\pm }\rangle _{a13}$\ and then convey the
measurement results to Bob via 2-bit classical communication \{00 to $|\pi
^{+}\rangle _{a13}$; 01 to $|\pi ^{-}\rangle _{a13}$; 10 to $|\mu
^{+}\rangle _{a13}$, 11 to $|\mu ^{-}\rangle _{a13}$\}. Then, via \{00 to $%
I; $ 01 to $\sigma _{3}$; 10 to $\sigma _{1}$; 11 to $\sigma _{1}\sigma _{3}$%
\} Bob can apply one of the local unitary operations \{$\mathit{I,}$ $\sigma
_{3},\sigma _{1},\sigma _{1}\sigma _{3}$\} to his qubit, then he converts
the state of his qubit `2' to that of the qubit `a'.

Similarly, a calculation yields

\begin{eqnarray*}
&&|\varphi \rangle _{a}|\eta ^{\prime }\rangle _{123} \\
&=&(\alpha |0\rangle _{a}+\beta |1\rangle _{a})|\eta ^{\prime }\rangle _{123}
\\
&=&\frac{1}{2}[|\pi ^{\prime +}\rangle _{a13}(\alpha |0\rangle _{2}+\beta
|1\rangle _{2}) \\
&&+|\pi ^{\prime -}\rangle _{a13}(\alpha |0\rangle _{2}-\beta |1\rangle _{2})
\\
&&+|\mu ^{\prime +}\rangle _{a13}(\beta |0\rangle _{2}+\alpha |1\rangle _{2})
\\
&&+|\mu ^{\prime -}\rangle _{a13}(\beta |0\rangle _{2}-\alpha |1\rangle
_{2})],
\end{eqnarray*}%
where
\begin{eqnarray*}
|\pi ^{\prime \pm }\rangle _{a13} &=&\frac{1}{2}|001\rangle _{a13}+\frac{1}{2%
}|010\rangle _{a13}\pm \frac{1}{\sqrt{2}}|100\rangle _{a13}, \\
|\mu ^{\prime \pm }\rangle _{a13} &=&\pm \frac{1}{\sqrt{2}}|000\rangle
_{a13}+\frac{1}{2}|101\rangle _{a13}+\frac{1}{2}|110\rangle _{a13}.
\end{eqnarray*}

Via the above discussion for $|\eta \rangle _{123}$, Alice can teleport the
unknown state $|\varphi \rangle _{a}$\ by sharing $|\eta ^{\prime }\rangle
_{123}$, where Alice has qubits `1' and `3' while Bob has the qubit `2'.

\section{Appendix B. For perfect superdense coding}

\setcounter{equation}{0} \renewcommand{\theequation}{B\arabic{equation}}

\subsection{ Perfect superdense coding by sharing $|\protect\eta \rangle $
and $|\protect\eta ^{\prime }\rangle $, respectively}

Let Alice and Bob share $|\eta \rangle _{123}$, where Alice has qubit 2
while Bob has qubits 1 and 3. Alice can apply unitary operators in \{$%
I,\sigma _{3},\sigma _{1},\sigma _{3}\sigma _{1}$\} on her qubit. That is,
Alice can apply the following local unitary operators $A_{i}^{(2)}$, $%
i=1,2,3,4$, \ to the state $|\eta \rangle $.

\begin{eqnarray}
A_{1}^{(2)} &=&I\otimes I\otimes I,  \label{loc-1} \\
A_{2}^{(2)} &=&I\otimes \sigma _{3}\otimes I,  \label{loc-2} \\
A_{3}^{(2)} &=&I\otimes \sigma _{1}\otimes I,  \label{loc-3} \\
A_{4}^{(2)} &=&I\otimes \sigma _{3}\sigma _{1}\otimes I.  \label{loc-4}
\end{eqnarray}

Thus, we obtain the following states $A_{i}^{(2)}|\eta \rangle $, $i=1,2,3,4$%
,

\begin{eqnarray}
A_{1}^{(2)}|\eta \rangle &=&\frac{1}{2}|000\rangle +\frac{1}{2}|101\rangle +%
\frac{1}{\sqrt{2}}|110\rangle , \\
A_{2}^{(2)}|\eta \rangle &=&\frac{1}{2}|000\rangle +\frac{1}{2}|101\rangle -%
\frac{1}{\sqrt{2}}|110\rangle , \\
A_{3}^{(2)}|\eta \rangle &=&\frac{1}{2}|010\rangle +\frac{1}{2}|111\rangle +%
\frac{1}{\sqrt{2}}|100\rangle , \\
A_{4}^{(2)}|\eta \rangle &=&-\frac{1}{2}|010\rangle -\frac{1}{2}|111\rangle +%
\frac{1}{\sqrt{2}}|100\rangle .
\end{eqnarray}

It is easy to check that $A_{i}^{(2)}|\eta \rangle $, $i=1,2,3,4$, are
mutually orthogonal. After Alice sends her qubit to Bob, Bob can make a
three-particle von-Neumann measurement on qubits 1, 2 and 3 using the four
mutually orthogonal states $A_{i}^{(2)}|\eta \rangle $, $i=1,2,3,4$. Thus,
Bob can determine what unitary operator Alice has applied. It means that $%
|\eta \rangle $ can be used to transmit 2-bit classical bits by sending one
qubit.

Similarly, Alice can apply the local unitary operators $A_{i}^{(2)}$, $%
i=1,2,3,4$, \ to the state $|\eta ^{\prime }\rangle $. Then, we obtain
\begin{eqnarray}
A_{1}^{(2)}|\eta ^{\prime }\rangle &=&\frac{1}{2}|001\rangle +\frac{1}{\sqrt{%
2}}|010\rangle +\frac{1}{2}|100\rangle , \\
A_{2}^{(2)}|\eta ^{\prime }\rangle &=&\frac{1}{2}|001\rangle -\frac{1}{\sqrt{%
2}}|010\rangle +\frac{1}{2}|100\rangle , \\
A_{3}^{(2)}|\eta ^{\prime }\rangle &=&\frac{1}{2}|011\rangle +\frac{1}{\sqrt{%
2}}|000\rangle +\frac{1}{2}|110\rangle , \\
A_{4}^{(2)}|\eta ^{\prime }\rangle &=&-\frac{1}{2}|011\rangle +\frac{1}{%
\sqrt{2}}|000\rangle -\frac{1}{2}|110\rangle .
\end{eqnarray}

Via the above discussion for $|\eta \rangle $, one can know that $|\eta
^{\prime }\rangle $ can be used to transmit 2-bit classical bits by sharing
the state $|\eta ^{\prime }\rangle $\ and sending one qubit, where Alice has
qubit 2 while Bob has qubits 1 and 3.

\subsection{Perfect superdense coding by sharing $|\protect\xi \rangle $ and
$|\protect\xi ^{\prime }\rangle $, respectively}

Let Alice and Bob share $|\xi \rangle _{123}$, where Alice has qubit 3 while
Bob has qubits 1 and 2. Alice can apply unitary operators in \{$I,\sigma
_{3},\sigma _{1},\sigma _{3}\sigma _{1}$\} to her qubit. That is, Alice can
apply the following local unitary operators $A_{i}^{(3)}$, $i=1,2,3,4$, \ to
the state $|\xi \rangle $.

\begin{center}
\begin{eqnarray}
A_{1}^{(3)} &=&I\otimes I\otimes I, \\
A_{2}^{(3)} &=&I\otimes I\otimes \sigma _{3}, \\
A_{3}^{(3)} &=&I\otimes I\otimes \sigma _{1}, \\
A_{4}^{(3)} &=&I\otimes I\otimes \sigma _{3}\sigma _{1}.
\end{eqnarray}
\end{center}

Thus, we obtain the following states $A_{i}^{(3)}|\xi \rangle $, $i=1,2,3,4$,

\begin{eqnarray}
A_{1}^{(3)}|\xi \rangle &=&\frac{1}{2}|000\rangle +\frac{1}{\sqrt{2}}%
|101\rangle +\frac{1}{2}|110\rangle , \\
A_{2}^{(3)}|\xi \rangle &=&\frac{1}{2}|000\rangle -\frac{1}{\sqrt{2}}%
|101\rangle +\frac{1}{2}|110\rangle , \\
A_{3}^{(3)}|\xi \rangle &=&\frac{1}{2}|001\rangle +\frac{1}{\sqrt{2}}%
|100\rangle +\frac{1}{2}|111\rangle , \\
A_{4}^{(3)}|\xi \rangle &=&-\frac{1}{2}|001\rangle +\frac{1}{\sqrt{2}}%
|100\rangle -\frac{1}{2}|111\rangle .
\end{eqnarray}

It is easy to check that $A_{i}^{(3)}|\xi \rangle $, $i=1,2,3,4$, are
mutually orthogonal. After Alice sends her qubit to Bob, Bob can make a
three-particle von-Neumann measurement on qubits 1, 2 and 3 using the four
mutually orthogonal states $A_{i}^{(3)}|\xi \rangle $, $i=1,2,3,4$. Thus,
Bob can determine what unitary operator Alice has applied. It means that $%
|\xi \rangle $ can be used to transmit 2-bit classical bits by sending one
qubit.

Similarly, Alice can apply the local unitary operators $A_{i}^{(3)}$, $%
i=1,2,3,4$, \ to the state $|\xi ^{\prime }\rangle $. Then, we obtain
\begin{eqnarray}
A_{1}^{(3)}|\xi ^{\prime }\rangle &=&\frac{1}{\sqrt{2}}|001\rangle +\frac{1}{%
2}|010\rangle +\frac{1}{2}|100\rangle , \\
A_{2}^{(3)}|\xi ^{\prime }\rangle &=&-\frac{1}{\sqrt{2}}|001\rangle +\frac{1%
}{2}|010\rangle +\frac{1}{2}|100\rangle , \\
A_{3}^{(3)}|\xi ^{\prime }\rangle &=&\frac{1}{\sqrt{2}}|000\rangle +\frac{1}{%
2}|011\rangle +\frac{1}{2}|101\rangle , \\
A_{4}^{(3)}|\xi ^{\prime }\rangle &=&\frac{1}{\sqrt{2}}|000\rangle -\frac{1}{%
2}|011\rangle -\frac{1}{2}|101\rangle .
\end{eqnarray}

Via the above discussion for $|\xi \rangle $, one can know that $|\xi
^{\prime }\rangle $ can be used to transmit 2-bit classical bits by sharing
the state $|\xi ^{\prime }\rangle $\ and\ sending one qubit, where Alice has
qubit 3 while Bob has qubits 1 and 2.

Statements and declarations: No financial interests, no competing interests,
no financial supports.

Data Availability Statement: No Data associated in the manuscript.

Acknowledgements

Thank the reviewers for the useful comments and Julia Huang (of Stanford
University) for changing English.

\end{document}